\documentstyle[sprocl,12pt]{article}

\begin{document}

\rightline{UTHEP-357}

\rightline{Jan. 1997}

\title{HADRONS IN DENSE MATTER\footnote{Invited talk
 at the 25th INS International Symposium on
 Nuclear and Particle Physics with High-Intensity Proton Accelerators,
 (Dec. 3-6, 1996, Tokyo, Japan.)}}

\author{Tetsuo Hatsuda}

\address{Institute of Physics, Univ. of Tsukuba, 
 Tsukuba, Ibaraki 305, Japan}

\maketitle

\abstracts{
 Physical reason behind the mass-shift of
 vector mesons  ($\rho$, $\omega$, $\phi$)
 in nuclear matter is discussed in the Walecka model and in QCD sum rules.
  Using  analytic  formulas for the mass-shift  valid at low densities,
  it is shown that the energy dependent part of the self-energy
  in medium (wave function renormalization)
  is  a main 
 source for the negative mass-shift, while  the energy independent
 part (the plasma frequency) 
 has  relatively a small effect.
  Future experiments in GSI ($\pi^- + A \rightarrow X + \omega$ )
 and KEK-PS ($p + A \rightarrow X + \phi$ ) to detect the twin peak
 structure of vector mesons in $e^+e^-$ spectrum are reviewed. Also, 
  it is proposed a new possible experiment using a  $S ({\rm strangeness}) = 
 1$ vector meson $K^{*}(892)$ with its radiative decay $K^{*+}(892)
  \rightarrow  K^+ + \gamma$.
}
  
\section{Introduction}

  Recently,  medium modification of hadrons acquires a lot of attention
 both in theories and experiments (see reviews \cite{original}.)
  Recent CERES data 
 showing a large enhancement of the
 $e^+ e^-$ pairs in central S+Au collisions with 200GeV/A
  give an experimental   hint for such medium effect and 
 induce theoretical studies \cite{2}.
 
  In this talk I will concentrate on the phenomena at zero temperature
 with finite baryon density, and discuss if there is a significant
  medium modification on
 the light vector mesons ($\rho, \omega, \phi, K^*$)
 in nuclear matter and nucleus.
 The answer is affirmative theoretically as will be explained below.
  I also discuss recent proposed
 experiments in GIS and KEK to detect such effects
 in hadron-nucleus reactions, and propose a new experiments using
 the radiative decay of $K^*(892)$.

\section{Vector Meson Poles in Medium}

 Let us first study general properties of the 
 vector meson propagator $D_{\mu \nu} $ in nuclear matter.
 For simplicity, I take a vector meson at rest ({\bf p}=0).
 Then, the longitudinal and transverse parts 
 become degenerate and the propagator  reduces to
\begin{eqnarray}
\label{prop}
D_{\mu \nu} (\omega^2) \propto {1 \over \omega^2 - m^2 - \Sigma(\omega^2)}\  ,
\end{eqnarray}
where $\omega$, $m$ and $\Sigma$ denote the frequency, the mass in the vacuum, 
 and the self energy in medium, respectively.
 The exact dispersion relation is obtained by $D^{-1}=0$ which gives
 a vector meson pole inside the medium as $\omega = m^*$.

 To extract more physics from eq.(\ref{prop}), let us derive an 
 approximate dispersion relation  by assuming
 that $\Sigma$ is a smooth function of $\omega$ near the origin:
 $\Sigma(\omega^2) \simeq A + B \omega^2$.
 Also, let us further assume that the density of the system is low enough
 so that $\Sigma$ can be treated as a perturbation.
  Whether the first assumption is valid or not is not  known in QCD
 and depends on models to evaluate $\Sigma$.  
 The second assumption is valid at least at sufficiently low density.
  Under these assumptions, one obtains an approximate dispersion relation
\begin{eqnarray}
\label{disp1}
(1-B)\  \omega^2 - m^2 - A =0,
\end{eqnarray}
which gives an approximate pole
\begin{eqnarray}
\label{disp2}
m^{*2} \simeq m^2 + A + m^2 B.
\end{eqnarray}
The sign and magnitude of $A$ and $B$ depend on the
 system one treats. For the photon propagation
 in a degenerate electron gas, $m=0$, $A=e^2 n_e /m_e > 0 $ and $B=0$.
 Namely, $m^*$ represents the well-known plasma frequency.
  On the other hand, in QCD,
 there is a possibility that $A > 0$ and  $B < 0$ with $A+m^2 B < 0$, i.e.
 the decreasing vector meson mass in medium.
 In the following sections, we will demonstrate this explicitly 
 in two different approaches (the Walecka model and QCD sum rules).

\section{Pole Shift in the Walecka Model}

 Let us take the $\omega$ meson at rest in nuclear matter.
 The on-shell properties of the $\omega$-meson in the Walecka model
 with vacuum polarization  were
 first studied by Saito, Maruyama and Soutome, and by 
 Kurasawa and Suzuki \cite{3}. Also, good physical arguments
 were given later by Jean, Piekarewicz and Williams \cite{4}.

 In the Walecka model,  the validity of the
 expansion $\Sigma(\omega^2) \simeq A + B \omega^2$
 can be checked explicitly, and the neglected terms are found to be 
 $O(m^2/4M^2) \simeq 15 \%$ or higher. ($M$ is the nucleon mass.)
   $A$ ($B$) in eq.(\ref{disp1})
 calculated within this approximation 
 comes from the coupling of the $\omega$-meson 
 to the particle-hole excitation
 (nucleon$-$anti-nucleon excitation). Then,
  the $\omega$-meson pole in nuclear matter at low density reads
\begin{eqnarray}
\label{disp3}
m^{*2} \simeq  m^2 + A \left[ 1 - 
 {4m^2 \over 3 \pi m_{\sigma}^2} {g_{\sigma}^2 \over 4 \pi} \right] ,
\end{eqnarray}
where $A = g_{\omega}^2 \rho /M$ (plasma frequency)
 with  $g_{\omega} (g_{\sigma})$ being  the $\omega (\sigma)$-nucleon
 vector (scalar) coupling constant,  and $m_{\sigma}$ being the
 $\sigma$-meson mass in the Walecka model. 
 Because of the large and negative contribution originating  from $B$
 (the term proportional to $g_{\sigma}^2$), the pole shift is negative, 
 $m^* - m < 0$.

 The sign of $A$ and $B$ can be understood easily by quantum mechanical
 level-repulsion due to the  second order perturbation.
 $A$ comes from the coupling of the $\omega$-meson to the 
 low-lying particle-hole continuum. Thus $A$ must be positive
 due to the level-repulsion.
 $B$ comes from the coupling of the $\omega$-meson 
 to the high-lying $N\bar{N}$ continuum. Since the 
 continuum threshold decreases because of the decreasing 
 nucleon's scalar-mass in the Walecka model, the coupling becomes stronger in 
 the medium than in the vacuum.  This causes negative $B$.

 Up to now, we have derived an approximate formula which is
 only valid when the system is in low densities  and simultaneously
 the expansion by $(m/2M)^2$ is valid.
 In ref.\cite{5},   one can see numerical results obtained
 by solving the full dispersion relation $\omega^2 - m^2 - \Sigma(\omega^2)
 =0$ for $\rho$ and $\omega$ mesons.
  On find that there is a considerable non-linearity,
  and the approximate formula (\ref{disp3}) is 
 valid only up to (0.2 $\sim $ 0.3)$\rho_0$.  Nevertheless, the
 physics extracted from eq.(\ref{disp3})  is qualitatively right.
 Generalization of this approach to the neutron matter and 
 asymmetric  nuclear matter has been also done \cite{dutt}.

\vspace{0.2cm}

Several comments  are in order here.

\begin{enumerate}

\item  Since $m^* < m$ is mainly caused by the 
 short distant $N \bar{N}$ excitation, one
  may derive an effective mesonic action by contracting the nucleon-loop
 into a point in the coordinate space.
 This gives an effective lagrangian
 $ {\cal L}_{eff} \propto \sigma^n F_{\mu \nu}^2 $ 
  with $\sigma$ being the scalar field and $F_{\mu \nu}$ being the
 field-strength tensor for the $\omega$-meson.
 Because of the baryon number conservation, one cannot generate
 an explicit mass term such as $\sigma^n \omega_{\mu}^2$
 from the nucleon loop.
 This feature may be considered to be  consistent with the phenomenological 
 analyses by Friman and Soyeur \cite{6} 
 on the near-threshold photo-production of $\rho$ and $\omega$ mesons.

\item  For the $\rho$-meson, one can imagine considerable collisional
 broadenings from the processes such as $\rho + N \rightarrow \Delta,\  
\Delta +\pi$. This has been checked, and it was shown that
 there is no significant effect from the collisional broadening near and below
 $\rho_0$ as far as the pole position $m^*$ decreases in medium \cite{7}.

\item Since the $N \bar{N}$ excitation only modifies the 
 wave function renormalization part of the propagator $B$,
 the behavior of the vector meson propagator has
 peculiar $\omega^2$ dependence.   For example,
 near the pole position,  it has a form
 $D(\omega^2 \sim m^{*2} ) \simeq A/(\omega^2 -
 m^{*2})$, while at $\omega^2=0$, it becomes
 $D(\omega^2 = 0) \simeq 1/(-m^2)$. Namely,
 the vacuum mass $m$ instead of $m^*$ appears near $\omega^2=0$.

\item 
 Unfortunately, the Walecka model is not a consistent effective field
 theory of QCD, since it has no expansion parameter which can control the
 higher dimensional operators of hadron fields unlike the
 case of the chiral perturbation theory.
 Also, the effect of the meson-baryon form factors in the 
 $N \bar{N}$ loop could largely attenuate the reduction of
  $m^*$.  Therefore, it is rather tempting to try more fundamental approaches
 to the problem. The QCD sum rules and the lattice QCD are the two 
 promising candidates.  The latter, however, has still
 problems to do simulations with finite chemical potential $\mu$ 
 although there may be a way out \cite{8}.  So, we will review the 
 results of the first approach in the next section.

\end{enumerate}

\section{Pole Shift in QCD Sum Rules}

The QCD sum rules (QSR) can be regarded as energy weighted sum rules
 which are familiar in atomic and nuclear physics.
  A consistent formulation of the 
 QSR in medium was first developed by Hatsuda and Lee \cite{9}.

 For the vector mesons, the starting point is a two point current correlation
 function in nuclear matter
\begin{eqnarray}
\Pi_{\mu \nu} (\omega^2) 
= i \int d^4x e^{i \omega t} \theta (t) \langle
 [\bar{q} \gamma_{\mu} q (x), \bar{q} \gamma_{\nu} q (0)] \rangle_{\rho},
\end{eqnarray}
with $\langle \cdot \rangle_{\rho}$ denotes an expectation value in 
nuclear matter with density $\rho$.
A set of sum rules (finite energy sum rules)
 can be derived from the above definition together 
 with the short distant  operator product expansion in QCD.
 The result is
\begin{eqnarray}
\label{sum}
\int_0^{\infty} ds\  s^n \  [{\rm Im} \Pi(s) - {\rm Im} \Pi_{pQCD}(s)]
 = C_n \langle {\cal O}_n \rangle_{\rho}.
\end{eqnarray}
Here $\Pi_{pQCD}(s)$ is a correlation function calculated 
 in perturbative QCD, $C_n$ is a Wilson coefficient and 
 ${\cal O}_n$ is a local operator such as
 $\bar{q}q$, $\bar{q} \gamma_{\mu} D_{\nu}q$, $(\bar{q}q)^2$.
 By making an ansatz 
 ${\rm Im}\Pi(s)= Z \delta (s-m^{*2}) + {\rm continuum}$, one
 can extract the pole position and residue in the medium 
 using the information on the matrix elements of local operators.

 Eq.(\ref{sum}) applied in the vacuum gives
 an approximate formula for the $\rho$ and $\omega$ meson masses \cite{10}
\begin{eqnarray}
m^2 \simeq [{448 \over 27} \pi^3 \alpha_s \langle \bar{q}q \rangle_0^2]^{1/3},
\end{eqnarray}
where relatively small contribution from the gluon condensate
 is neglected. 
In the low density medium, one can derive an approximate  formula
 $m^{*2} \simeq m^2 + A + m^2 B $ with
\begin{eqnarray}
\label{AB1}
A & = & 2 \pi^2 {\rho \over M}
 \left( 1- \bar{x} {M^2 \over m^2} \right) > 0 ,\\
\label{AB2}
m^2(1+B) & = & 
 m^2 \left( { {\langle \bar{q} q \rangle_{\rho}} \over 
        {\langle \bar{q} q \rangle_{0}}        }\right) ^{2/3} 
 - 4 \pi^2 {\rho \over M_N} \bar{x} {M_N^2 \over m^2} < m^2 ,
\end{eqnarray}
where the first (second)  term in the right hand side of
 eq.(\ref{AB2})  comes from the 
 4-quark condensate (the twist 2 matrix element).
 $\bar{x}$ denotes the quark momentum fraction in the nucleon
  which is measured in the deep inelastic scattering.
 The gluon condensate, mixed quark-gluon condensate and
 a dimension 6 condensate with twist 2 are neglected in the
 above formula. Also, 
 a factorization assumption for the 4-quark condensate in medium
 is adopted.

 Because of the smallness of the plasmon-like term $A > 0$ compared
 to $B <0$, $m^*$
 generally decreases in nuclear  matter.
 Also, the formula without A and $\bar{x}$ has a similar structure
 with that predicted by Brown and Rho \cite{BR} where KSRF relation is
 assumed to be valid in medium. 
 .
 
 The formulas eq.(\ref{AB1},\ref{AB2}) are valid only at low densities, and 
 one should solve the full sum rules eq.(\ref{sum}) numerically to get
 quantitative predictions.
 The key parameter for such calculation is the ratio
 of the quark condensate in matter and that in the 
 vacuum. The exact expression
 for such ratio in QCD reads \cite{11}
\begin{eqnarray}
 {{\langle \bar{q} q \rangle_{\rho}} \over 
        {\langle \bar{q} q \rangle_{0}}        }
 = 1- {\rho \over f_{\pi}^2 m_{\pi}^2} \left[ \Sigma_{\pi N}
 + m {d \over dm} \left( E(\rho )/A \right)  \right] .
\end{eqnarray}
Here $\Sigma_{\pi N} =45 \pm 10$ MeV is the pion-nucleon sigma term,
 and $E/A$ is a nuclear binding energy per particle with $m$ being the
 current quark mass.  The nuclear binding effect is known to be
 rather small around and below $\rho_0$ \cite{12}.
  Using the results of 
 numerical calculation and fitting  them  by a linear form, one obtains
\begin{eqnarray}
\label{ms1}
{m_{\rho,\omega}^* \over m_{\rho,\omega}} =  1- (0.16 \pm 0.06)\  
{\rho \over \rho_0}, \ \ \ \ \ \ 
{m_{\phi}^* \over m_{\phi}} =  1- (0.023 \pm 0.011)\   
{\rho \over \rho_0}.
\end{eqnarray}
 The errors in the above formulas are originating from the uncertainties
 of the density dependent  condensates.
 The contribution of the
  quark-gluon mixed operator with twist 4,  which may
   possibly weaken the mass shift \cite{9}, is neglected in the above.
 Also the validity of the factorization assumption for the 4-quark condensate
 used in the above should be checked by other method.

 I should emphasize here that the formulas eq.(\ref{ms1})
 does not imply that the density dependence of $m^*$ is 
 strictly linear.  The actual numerical results in \cite{9,14}
 show non-negligible non-linearity of $m^*(\rho)/m$
  below $\rho_0$.

\vspace{0.2cm}

Some comments are in order here.

\begin{enumerate}

\item Asakawa and Ko have introduced  a realistic
 spectral function   by taking into account
 the width of the $\rho$-meson and the effect of the collisional
 broadening due to the   $\pi$-$N$-$\Delta$-$\rho$
 dynamics \cite{13}.  By the similar QSR analyses as above,
 they found that the negative mass shift 
  persists even in this realistic case.  The width of the
 rho meson in their calculation  decreases as density increases, which 
implies that the phase space suppression from the $\rho \rightarrow 2 \pi$
 process overcomes the collisional broadening at finite density.
 Further examination of this interplay between the mass shift and
 the collisional broadening for the $\rho$-meson
 is important in relation to the future
 experiments.

\item Monte Calro based error analysis was applied to QSR
  by Jin and Leinweber \cite{14} in place of the
 Borel stability  analysis employed in \cite{9}.
 They found 
 $ m_{\rho,\omega}^*/ m_{\rho,\omega} =  1- (0.22 \pm 0.08)  
(\rho /\rho_0)$ and 
 $m_{\phi}^* /m_{\phi} =  1- (0.01 \pm 0.01) 
(\rho / \rho_0)$, which are consistent with
 eqs. (\ref{ms1}) within 
 the error bars.

\item  Very recently Koike and Hayashigaki  re-analyzed the 
    effective scattering amplitude $\bar{f}_{VN}$
 defined by 
$  m^{*2} \simeq m^2 + {\bar f}_{VN} \cdot \rho\ \ $
  using the QSR in the vacuum \cite{16}. 
 This  analysis shows that (a) the
  previous calculation by the same author \cite{koi}
 is incomplete (as has been already pointed out  in \cite{9,14}),
  and (b) a negative scattering length is obtained; ${\bar f}_{VN} < 0$.
 The latter feature  supports the decreasing vector
 meson masses.  The decrease obtained in 
 \cite{16} for $\rho$ and $\omega$ mesons are factor 2 smaller than
 that in \cite{9}. This is party because the Borel 
 stability of the scattering
 length calculation is not as excellent as that of the  QSR in medium, and
 partly because the scattering-length approach gives 
 a formula only valid at extremely low density. 
 In fact the number obtained in \cite{16} is consistent with
  that of the numerical result of $m^*(\rho)/m$ near $\rho=0$
  given in Fig.2(a) of \cite{9} and in Fig.1 of \cite{14}.

\item   Jaminon and Ripka has reached a similar pole shift
   by using a model of vector mesons coupled to  constituent quarks
  \cite{17}.
   Saito and Thomas have examined a  different 
 but comprehensive model (bag model combined with the Walecka model)
 and 
 found  decreasing vector-meson masses \cite{18};
 $ m_{\rho,\omega}^* / m_{\rho,\omega}  \sim   1- 0.09  
(\rho / \rho_0)$.
  The spectral shift 
 of the quarks inside the bag induced by the existence of 
 nuclear medium plays a key role in this approach. 

\item The three momentum $\vec{p}$ dependence of the dispersion relation
  $\omega^2 = m^{*2} + (1+a) \vec{p}^2$
 of the vector-meson in QSR
  has been also studied recently by Lee and Friman \cite{181}. 
 They found that $ \mid a \mid < 0.08$ at nuclear matter density.
  Walecka model also predicts small $a$. \cite{182}

\item Eletskii and Ioffe has analysed an ``effective mass'' of
 the rho-meson passing through the nuclear matter \cite{183}. Since their
 approach is limited only to the fast rho-mesons having the kinetic
 energy more than 2 GeV, it does not have direct relevance to 
 the physics discussed in this article.

\end{enumerate}

   Basic idea common in the approaches predicting the
 decreasing meson mass (at rest) may be summarized as follows.
 In nuclear matter, scalar ($\sigma$) and vector ($\omega$)
 mean-fields are induced by the  nucleon sources.
  These mean-fields give back-reactions
 to the nucleon propagation in nuclear matter and modify
  its self-energy. This is an origin of the effective nucleon mass
 $M^* < M$ in  the relativistic
 models for nuclear matter.
  The same mean-fields should also affect the propagation of 
  vector mesons in nuclear medium.
 In QSR, the quark condensates act on the quark propagator
 as density dependent mean-fields.
  In the Walecka model, the coupling of the mean-field with the vector
 mesons are taken into account through the short distant
  nucleon
  loop with the effective mass $M^*$.
 An interesting observation is that major part of the
 mean-field contributes to modify the wave function renormalization
 constant $B$  as shown in (\ref{disp3})
 and in (\ref{AB2}).

\section{Possible Experiments}

How one can detect the spectral change of vector mesons  in experiments?
 As have mentioned in the Introduction,
 enhancement of the lepton pairs below the $\rho$-resonance region
 in  S+Au collisions 
 was reported by CERES/NA45 at CERN \cite{19}.
 Similar
 enhancement of the muon pairs is  reported in S+W collisions
 by HELIOS-3 at CERN too \cite{20}.  This enhancement is rather
 difficult to explain by conventional mechanisms of the lepton
 pair production such as the Dalitz decay, $\pi^+ \pi^-$ annihilation
 and the $\rho$-decay \cite{2}. Although the assumption
 of the  decreasing $\rho$-mass can explain the data well \cite{2},
  it is not an unambiguous proof of the
 mass shift because complicated dynamics of the heavy-ion 
 collisions are involved in the data analyses.

  In higher energy heavy ion collisions such as 
 RHIC and LHC, high temperature plasma 
 with low baryon density will be formed in the central region.
 In this case, the twin peak structure of the $\phi$-meson proposed by
  Asakawa and Ko \cite{21} is a very interesting and clean
 signature of the mass shift.

 On the other hand, around the normal nuclear matter density at
zero temperature, one could see the mass shift 
 in various hadronic or electromagnetic production of the
 vector mesons with heavy nuclear target. A typical 
 signal of the mass shift in these cases is  the twin
 peak structure similar to the one that Asakawa and Ko proposed.

Suppose that one creats the vector meson inside the 
 nucleus by $\pi, K, \gamma$ or $p$ beams.
 The total number of lepton pairs from the
 decay of vector mesons inside the nucleus is
 roughly estimated as 
\begin{eqnarray}
\label{br}
N_{in}(e^+e^-) \simeq
 N \times (1- e^{-\Gamma_{tot}^* R}) \times {\rm Br}(e^+e^-),
\end{eqnarray}
where $N$ is the total number of created vector mesons, 
 $\Gamma_{tot}$ is the total width of the vector meson in the nucleus,
 $R$ is the nuclear radius, and  ${\rm Br}(e^+e^-)$ is a branching ratio
 to the $e^+ e^-$ decay.
 The second factor in the right hand side of eq.(\ref{br}) is a probability 
 to have vector mesons decaying inside.

\vspace{0.2cm}
 
 Some comments are in order here.

\begin{enumerate}

\item  One can {\em effectively} increase the 
 number of vector mesons decaying inside the nucleus
  by choosing a kinematics of producing
 ``recoilless'' or ``stopped'' vector mesons.

\item  The particle width in nuclear matter $\Gamma_{tot}^* $ can be  
 quite different from the width in the vacuum $\Gamma_{tot}$ 
  given in Table 1.
 The collisional broadening increases the total-width, while the
 decreasing $m^*$ tends to make the total-width small due to the 
 phase space suppression. Unfortunatelly,  we do not know exactly 
 how $\Gamma_{tot}^*$ behaves as a function of nuclear density.
 If $\Gamma_{tot}^* \sim \Gamma_{tot}$ (which may be quite wrong),
  most of the $\rho$ ($\phi$) mesons  decay inside (outside)
  the nucleus. On the other hand,
  because of its large (small) width, the invariant mass spectrum
 of lepton pairs is broad (sharp).
  The situation for $\omega$ meson is 
 just in between $\rho$ and $\phi$.

 \item  To get clean signals with small final state interactions, the
 detection of the lepton pair is the better than $\pi^+ \pi^-$
 or $K^+ K^-$, although the
 branching ratio is as small as  $10^{-4} \sim 10^{-5}$.
 Despite the strong final state interactions, 
 the radiative decays and hadronic decays of the
 vector mesons can be also used as signals.  

 \item
  $K^{*+} (892)$, which is a $S ({\rm strangeness})=  1$ vector meson, 
  does not decay into lepton pairs but decays into 
 $K^+ \gamma$ with sizable branching ratio of $10^{-3}$ (see Table 1).
 This meson has several advantages: (a) The total width  
 is 50 MeV which
 is sufficiently  large for $K^*$ to 
 decay in heavy nuclei and is sufficiently small  to get
 clean signal in  $K^+ \gamma$ spectrum.  (b) The branching ratio
 of $K^*$ 
 to $K^+ \gamma$ is order of magnitude larger than the leptonic branching
 ratios of neutral vector mesons. (c)  Because the final product
   $K^+$ has quark
 composition $\bar{s}u$, $K^+$ 
  has a long mean free path
 in nuclear matter ( 5$\sim$6 fm) and does not suffer  final
 state interactions so much.
  Thus, $K^{*+} (892)$ 
 supplies a new possibility of detecting the mass shift, 
  which has not been addressed so far. Studies of pole shift
 of $K^*$ in the Walecka model and in QCD sum rules as well as
 its detectability in heavy-ion collisions and hadron-nucleus
 collisions are now under way \cite{100}.

\end{enumerate}

\begin{table}[t]  
\caption{
 Vector Mesons below 1 GeV and its total width, leptonic 
 branching ratio, radiative $K \gamma$ branching ratio,
 and proposed experiments.}
\vspace{0.4cm}
\begin{center}
\begin{tabular}{|c|c|c|c|c|}
\hline
particle & $\Gamma_{tot}$  
  & Br($e^+e^-$)   &  Br($K \gamma$)  & 
 Proposed experiment \\ \hline \hline
 $\rho^0$ (770)  &  (1.3fm)$^{-1}$  &  $4.5 \times 10^{-5}$
& $--$   &   Spring-8  \\ 
 $\bar{u}u - \bar{d}d$ &  ($\pi \pi $)  &  & & $\gamma$+$A$ \ 
 ($E_{\gamma}<$2.5GeV) \\ \hline
 $\omega$ (782)  &  (23.5fm)$^{-1}$  &  $7.2 \times 10^{-5}$
& $--$   &   HADES at GSI  \\ 
 $\bar{u}u + \bar{d}d$ &  ($\pi \pi \pi $) &  & &
 $\pi^-$+$A$\  ($p_{\pi} \sim 
1.3$GeV/$c$) \\ \hline
 $\phi$ (1020)  &  (45fm)$^{-1}$  &  $3 \times 10^{-4}$
& $--$   &   E325 at KEK-PS  \\ 
 $\bar{s}s$ &  ($K K $)  &  & & $p$+$A$ \  ($E_{p}\sim $12GeV) \\ \hline
 $K^*$ (892)  &  (3.9fm)$^{-1}$  &  0  
& $1 \times 10^{-3}$   &     \\ 
 e.g. $\bar{s}u$ &  ($K \pi $) &  & $K^{*+}$ 
$\rightarrow$$K^+$$\gamma $  &   \\ \hline  
 \end{tabular}
\end{center}     
\end{table}

There is a proposal of detecting the $e^+e^-$ pairs
 from the reaction 
 $\pi^- + A \rightarrow X + \omega \ $
 using
  HADES at GSI \cite{22}.  The typical momentum of the incident  
 $\pi^-$ is 1.3 GeV/c, which can create
 substantial number of ``almost'' recoilless
  $\omega$ mesons inside the nucleus 
 (e.g. $p_{\omega} < 0.4 {\rm GeV/c}$). 
  This will give rise to a distinct twin $\omega$ peak structure
 as well as shifted broad $\rho$-peak in the lepton pair 
 spectrunm.

 In E325 experiment at KEK \cite{23}, 
the reaction 
 $p + A \rightarrow X + \phi\ $
  are used and $e^+e^-$ as well as $K^+K^-$
  will be measured.  The incident proton energy
 is 12 GeV which gives the typical $\phi$-meson momentum  
 1 GeV/$c$.  Still, one can see a twin peak structure in heavy
 nuclei:
 the  higher peak is the $\phi$ decaying outside and
 the lower peak is from the $\phi$ decaying inside.
 The change of the leptonic vs hadronic branching ratio
$r = \Gamma(\phi \rightarrow e^+ e^- ) /\Gamma (\phi \rightarrow K^+ K^- )$
 can be also measured.
 Since $m_{\phi}$ is very close to $2m_{K}$ in the vacuum,
 any modification of the $\phi$-mass or the $K$-mass changes
 the ratio $r$ substantially as a function of mass number of 
 the target nucleus.

 One can also do the similar experiments  in
 SPRING-8 in Japan using $\gamma + A $ reactions \cite{24}.

\section{Concluding Remarks}

The spectral change of the elementary excitations in medium
 is an exciting new possibility in QCD.  
 By studying such phenomenon, one can learn the structure
 of the hadrons and the QCD ground state at finite $(T, \rho$)
 simultaneously.   Theoretical approaches such as the QCD sum rules
 and the hadronic effective theories 
 predict that the  light vector mesons ($\rho$, $\omega$, $\phi$, $K^*$) 
 are sensitive to the partial restoration of chiral symmetry
 in hot/dense medium.  These mesons are good 
 probes experimentally, since they decay into lepton pairs which penetrate
 the hadronic medium without loosing much information.
  Thus, the lepton pair spectroscopy in QCD will tell us a 
 lot about the detailed structure of the hot/dense matter, which
 is quite similar to the soft-mode spectroscopy
 by the photon and neutron scattering experiments in solid state physics.

\section*{Acknowledgments}

This work was supported by the Grants-in-Aids of the Japanese Ministry
 of Education, Science and Culture (No. 06102004).

\end{document}